# Nanoscale probing of image-potential states and electron transfer doping in borophene polymorphs

Xiaolong Liu,[1] Luqing Wang,[2] Boris I. Yakobson,[2,3] and Mark C. Hersam[1,4,5,6*]

[1]Applied Physics Graduate Program, Northwestern University, Evanston, IL 60208, USA

[2]Department of Materials Science and NanoEngineering, Rice University, Houston, TX 77005, USA

[3]Department of Chemistry, Rice University, Houston, TX 77005, USA

[4]Department of Materials Science and Engineering, Northwestern University, Evanston, IL 60208, USA

[5]Department of Chemistry, Northwestern University, Evanston, IL 60208, USA

[6]Department of Electrical and Computer Engineering, Northwestern University, Evanston, IL 60208, USA

*Correspondence should be addressed to: m-hersam@northwestern.edu

**Abstract**

Using field-emission resonance spectroscopy with an ultrahigh vacuum scanning tunneling microscope, we reveal Stark-shifted image-potential states of the $v_{1/6}$ and $v_{1/5}$ borophene polymorphs on Ag(111) with long lifetimes, suggesting high borophene lattice and interface quality. These image-potential states allow the local work function and interfacial charge transfer of borophene to be probed at the nanoscale and test the widely employed self-doping model of borophene. Supported by apparent barrier height measurements and density functional theory calculations, electron transfer doping occurs for both borophene phases from the Ag(111) substrate. In contradiction with the self-doping model, a higher electron transfer doping level occurs for denser $v_{1/6}$ borophene compared to $v_{1/5}$ borophene, thus revealing the importance of substrate effects on borophene electron transfer.

**Main Text**

Entirely synthetic two-dimensional (2D) materials [1] have no bulk counterparts and are frequently studied with *in situ* techniques that characterize electronic structure in the vicinity of the Fermi level [2]. Image potential states (IPSs) [3], however, are high-lying quantized states in a Rydberg series just below the vacuum level [4], offering a sensitive probe of material quality and interfacial coupling in substrate-supported 2D materials. In this context, interfacial charge transfer across weakly interacting and chemically hybridized interfaces are expected to be vastly different [5–7], thus motivating the study of IPSs in synthetic 2D materials. Although interfacial coupling is of particular interest for synthetic 2D materials due to their stabilization by substrate interactions and/or buckling [8–12], IPSs have thus far only been observed in graphene and hexagonal boron nitride [13,14].

Borophene, the 2D synthetic form of boron, is metallic in all synthesized polymorphs [10,15–18] and hosts massless Dirac fermions [19], charge density waves [16], and potentially superconductivity [20]. Compared to carbon, the electron deficiency of boron does not favor a two-center honeycomb lattice because the strong in-plane $sp^2$ bonding states would otherwise be partially filled. This electron deficiency is mitigated by partial filling of hollow hexagons (HHs) with boron atoms [21–24], leading to energetically similar borophene polymorphs with periodic HHs (with a vacancy concentration of $v$) embedded in an otherwise triangular lattice. Their stability is explained by a self-doping picture that describes added boron atoms as pure donors providing 3 electrons/atom to a fixed electronic structure of the honeycomb lattice ($v = 1/3$) [22,23]. The preferred borophene phase thus obeys a rule of filling, which requires all in-plane $sp^2$ bonding states to be filled and isoelectronic with graphene. In vacuum, $v = 1/9$ achieves the optimal filling, while on substrates, any interfacial charge transfer alters the ground state structure accordingly – a phase with larger $v$ would be associated with greater electron transfer from the substrate. The self-doping model has offered a general computational design rule for a variety of boron nanomaterials including metal borides [22,25,26], boron hydrides [27], and boron fullerenes [21,28]. It is also the underlying mechanism of the proposed phase-tuning of borophene by making contact with electrides [29]. Despite its conceptual simplicity and wide applications, this self-doping model has not been verified experimentally since its proposal more than a decade ago. Computationally, the self-doping model alone has been shown to be insufficient to guarantee the existence of certain boron fullerenes [30]. In addition, the manner in which the generally believed stronger interfacial coupling (i.e., chemical hybridization) between synthetic 2D materials and substrates affects the filling rule is unclear. The answer to this outstanding question is highly relevant for the interface physics of borophene and determines the predictive power and practical utility of the self-doping model.

Herein, to address the knowledge gaps discussed above, we use an ultrahigh vacuum (UHV) scanning tunneling microscope (STM) at 4.2 K to perform FER spectroscopy, which reveals up to twelve IPSs of two borophene polymorphs ($v = 1/5, 1/6$). By examining local work functions (LWFs), we discovered a higher electron doping level in $v_{1/6}$ borophene compared to that of $v_{1/5}$ borophene, which is well-supported by density functional theory (DFT) calculations. However, this result contradicts the predictions of the self-doping model, which implies its limitations for substrate-supported borophene and highlights the importance of interfacial interactions in synthetic 2D materials involving chemical hybridization.

Simultaneous growth of $v_{1/5}$ and $v_{1/6}$ borophene on Ag(111) allows [16,18] reliable comparison of these two polymorphs with unchanged STM tip conditions. In an STM, when a positive sample bias exceeds the sample LWF, electrons tunnel in the Fowler-Nordheim regime. The high electric field in the junction causes a Stark shift of the hydrogenic spectrum of IPSs (Fig. 1(a)). Resonant tunneling occurs when the Fermi level of the tip aligns with one of these Stark-shifted IPSs (pink lines) [31]:

$$eV_n = \phi + \left(\frac{3n\pi\hbar eE}{2\sqrt{2m}}\right)^{\frac{2}{3}} \tag{1}$$

where $V_n$ is the sample voltage for the $n^{\text{th}}$ resonance, $\phi$ is the sample LWF, and $E$ is the electric field. Experimentally, the tunneling current during a bias sweep is maintained at a constant value while the tip height $z$ adjusts. In Fig. 1(b), the measured $z - V$ curves (vertically offset) on Ag(111), $v_{1/6}$ borophene, and $v_{1/5}$ borophene at different setpoint conditions exhibit a series of steps at bias voltages corresponding to $V_n$. The resonance at each IPS up to $n = 12$ is manifested as peaks in differential tunneling conductance curves ($\frac{dI}{dV} - V$) as shown in Fig. 1(c) as well as in $\frac{dz}{dV} - V$ curves as shown in Fig. S1, but is absent when the sample is biased negatively (Fig. S2). The peaks around 2 V on borophene are not IPSs (too low for FER) and thus are likely to be interface states [32]. For all three materials, as the current setpoint is increased, the total number of peaks in the FER spectrum decreases because a smaller tip-sample distance results in a stronger electric field that increases the energy separation between IPSs (equation (1)).

Due to scattering processes, the IPSs have finite lifetimes that can be calculated using $\tau_n = \hbar/\Gamma_n$ as shown in Fig. 1(d), where $\Gamma_n$ is the full width at half maximum of the $n^{\text{th}}$ IPS. The lifetimes of both borophene phases and Ag(111) increase with $n$, ultimately reaching values of 9 ps and 7 ps, respectively. This increasing lifetime can be explained by a reduced overlap of higher order IPS wavefunctions with the bulk of the substrate [33]. However, as suggested by the black lines, the increase of $\tau_n$ is close to linear for Ag(111) and sub-linear for borophene, in contrast to the theoretical $\tau_n \propto n^3$ relationship for field-free IPSs [33]. This deviation is a result of the tunneling junction electric field, which lifts higher-order IPSs up in energy and increases the number of available inelastic scattering channels. Furthermore, the electric field pushes the IPSs closer to the surface, enhancing scattering efficiency [34]. This effect of the electric field is also evident from the overall decreasing lifetimes with increasing tunneling current setpoints in all three materials. The fact that both borophene polymorphs possess similar lifetimes that are higher than Ag(111) is an indicator of the high quality of the borophene crystals with minimal in-plane corrugation, which would otherwise produce shorter-lived IPSs as observed for rippled graphene on Ru(0001) [35].

According to equation (1), IPSs allow quantitative determination of the LWF. A comparison of the $\frac{dI}{dV} - V$ curves on borophene and Ag(111) at the same current setpoint is shown in Fig. 2(a). The peak energies at each quantum number $n$ on $v_{1/5}$ borophene are consistently smaller than those of $v_{1/6}$ borophene, suggesting a lower LWF, but are significantly larger than those of Ag(111). The sharp transition of the IPSs from Ag(111) to $v_{1/6}$ borophene over ~2 nm in Fig. 2(b) suggests that it is possible to use IPSs as a nanoscale LWF probe. As shown in Fig. 2(c), the differential conductance map $g(\boldsymbol{r}, V)$ near the $n = 1$ IPS of borophene (~4.57 V) reveals nanoscale details such as the presence of subsurface defects that are not detected in the topographic image $T(\boldsymbol{r})$ or $g(\boldsymbol{r}, V)$ images at lower bias voltages (Fig. S3). Because the lower-order IPSs have a significant portion of their wavefunctions near the sample surface with resulting in-plane scattering [36,37], substrate defects result in broad and split $n = 1$ peaks as shown in Fig. 2(d). The green (blue) and red (yellow) spectra are taken on (off) the subsurface defects in Ag(111) and $v_{1/6}$ borophene, respectively, as indicated by the colored crosses in Fig. 2(c). By plotting $V_n$ of the $n^{\text{th}}$ IPS with respect to $n^{2/3}$ in Fig. 2(e), the data points lie along straight lines in agreement with equation (1). The LWFs are therefore determined by the y-axis intercepts of linear fits to $V_n - n^{2/3}$ data points. For better accuracy, only $n \geq 3$ IPSs are used for fitting. The results are given in Fig. 2(f). On average, the LWFs measured are: $\phi_{Ag} = 4.43 \pm 0.14$ eV, $\phi_{\frac{1}{6}/Ag} = 4.69 \pm 0.13$ eV, and $\phi_{\frac{1}{5}/Ag} = 4.64 \pm 0.13$ eV. The relative values of the LWFs are well supported by DFT calculations ($\phi_{Ag}^{DFT} = 4.43$ eV, $\phi_{\frac{1}{6}/Ag}^{DFT} = 4.64$ eV, and $\phi_{\frac{1}{5}/Ag}^{DFT} = 4.53$ eV) shown in Fig. 2(f). Furthermore, the measured and calculated Ag(111) LWF (4.43 eV) matches well with literature reports (4.46 eV) [38]. Since the surface potential of atomically thin borophene on Ag(111) is a convolution of that from borophene and Ag(111), the measured LWF should be considered as from the combined material system. The fact that both borophene phases show larger LWFs than Ag(111) suggests the formation of inward dipoles at the borophene/Ag(111) interface caused by interfacial electron transfer to borophene, increasing the total surface potential barrier for bulk electrons to escape into vacuum. From another perspective, the LWFs of both borophene phases on Ag(111) are smaller than those of free-standing borophene in vacuum (both ~4.9 eV), which is consistent with increased borophene chemical potentials due to electron transfer doping. Such interfacial electron transfer is not surprising and consistent with the self-doping picture, where electrons donated by the Ag(111) substrate stabilize $v_{1/5}$ and $v_{1/6}$ borophene instead of $v_{1/9}$ borophene. This substrate interaction can also explain the absence of a double Rydberg series of even and odd symmetry IPSs in borophene even though borophene is a 2D solid. A similar situation is encountered for graphene, where the double Rydberg series is only detected when interfacial coupling is exceptionally weak [39].

To independently test the LWFs extracted from IPSs, we also measured the apparent barrier heights (ABHs) by monitoring the tunneling current ($I$) as the tip is displaced in the normal direction of the surface. To first order, the logarithm of $I$ is linearly proportional to tip-sample distance $z$:

$$\ln I(z) \propto -2\sqrt{\frac{2m\phi_a}{\hbar^2}} z \qquad (2)$$

where $m$ is the electron mass, and $\phi_a$ is the ABH. The results shown in Fig. 3(a) clearly suggest a larger ABH of $v_{1/6}$ borophene (4.008 $\pm$ 0.004 eV) than $v_{1/5}$ borophene (3.882 $\pm$ 0.004 eV), both of which are significantly larger than that of Ag(111) (3.186 $\pm$ 0.003 eV). Although the tip

LWF and the exact tunneling barrier shape is unknown, these results are in agreement with the LWFs determined from IPSs and DFT calculations because ABHs correlate positively with sample LWFs [32]. By modulating the tip-sample distance $z$ during constant-current imaging and recording $|\partial lnI(\mathbf{r},z)/\partial z|$, the relative LWF across the boundary between $v_{1/5}$ borophene and Ag(111) can be directly visualized as demonstrated in Fig. 3(b). Although the borophene island covers adjacent Ag terraces shown in the topographic image ($T(\mathbf{r})$), the simultaneously acquired $|\partial lnI(\mathbf{r},z)/\partial z|$ image consistently reveals a larger LWF of $v_{1/5}$ borophene. The lower LWFs at three Ag step edges (red arrows in $T(\mathbf{r})$) are expected from 'Smoluchowski smoothing' [40], further supporting the validity of these measurements. A direct comparison between the ABHs of adjacent $v_{1/6}$ and $v_{1/5}$ borophene polymorphs shown in Fig. 3(c) again yields consistent results, confirming a higher electron transfer doping level in $v_{1/6}$ borophene than $v_{1/5}$ borophene. Compared with conventional topographic images, $|\partial lnI(\mathbf{r},z)/\partial z|$ imaging gives rise to improved atomic resolution (Fig. S4). Furthermore, the measured ABHs only significantly decrease when the tip makes point contact with the sample (Fig. S5).

To computationally compare the electron doping levels in the two borophene polymorphs, we performed Bader analysis of the charge distributions in relaxed $v_{1/5}$ and $v_{1/6}$ borophene on Ag(111). The isosurfaces of electron accumulation (blue, 0.002 e/Å$^3$) and depletion (red, -0.002 e/Å$^3$) of the $v_{1/6}$/Ag and $v_{1/5}$/Ag interfaces are shown in Figs. 4(a) and 4(b), respectively, which confirm overall electron transfer from the Ag(111) substrate to the borophene polymorphs. The in-plane charge distributions for $v_{1/6}$ borophene and $v_{1/5}$ borophene on Ag(111) are shown in Figs. 4(c) and 4(d), respectively. The areal charge accumulations are 0.0101 e/Å$^2$ in $v_{1/6}$ borophene and 0.0086 e/Å$^2$ in $v_{1/5}$ borophene (while the borophene-Ag spacing 2.375 Å is practically identical), in agreement with the LWF measurements. Therefore, both experiment and theory contradict the self-doping picture, which suggests more electron transfer from the substrate for a phase with a higher HH density (i.e., $v_{1/5}$). This result suggests that the filling rule of the self-doping model is not obeyed in either or both borophene polymorphs and/or the charges from self-doping and external doping are not simply addable and thus not controlling the phase of borophene, which further reveals the limitations of the self-doping model. Both explanations are likely due to the presence of interfacial chemical hybridization between borophene and the substrate that can alter the amount and even the sign of charge transfer across 2D material/metal interfaces compared to those with weaker interactions [5–7]. The strong presence of charges in the borophene/Ag gap in Fig. 4(a,b) supports formation of hybridized bonding orbitals, which likely alter the in-plane bonding configurations of borophene and the stabilization requirement of being isoelectronic with graphene. This situation is similar to chemisorbed graphene on metal surfaces, where the interfacial behavior and bonding configuration are dramatically different from physisorbed graphene such that simple charge transfer models are not applicable [7]. Under the influence of chemical hybridization, $v_{1/6}$ borophene becomes the most stable phase on Ag(111) substrates [5]. Because larger interfacial charge transfer leads to larger interfacial binding, it is thus not surprising that the substrate electron transfer doping level is higher for $v_{1/6}$ borophene. Indeed, the DFT calculated binding energy of $v_{1/6}$ borophene on Ag(111) is 0.048 eV/ Å$^2$ compared to 0.044 eV/ Å$^2$ for $v_{1/5}$ borophene. Here, the binding energy is defined as $E_{bind} = E_B + E_{Ag} - E_{B/Ag}$, where $E_{B/Ag}$, $E_B$, and $E_{Ag}$ are the total energies of borophene on Ag(111), freestanding borophene, and bare Ag(111) substrates, respectively. This analysis also suggests that the difference of the LWFs for $v_{1/6}$ and $v_{1/5}$ borophene contributes to the driving force of the observed periodic self-assembly of $v_{1/6}$ and $v_{1/5}$ borophene superlattices [16].

In summary, we have characterized the IPSs of $v_{1/6}$ and $v_{1/5}$ borophene polymorphs on Ag(111) using cryogenic UHV STM. Using IPS and ABH measurements and DFT calculations, we determined a higher substrate electron transfer doping for $v_{1/6}$ borophene compared to $v_{1/5}$ borophene, which contradicts the self-doping picture and highlights the importance of interfacial interactions in substrate-stabilized synthetic 2D materials. Overall, this work demonstrates that IPSs are a sensitive probe for studying interfacial interactions in borophene polymorphs, which can likely be generalized to other emerging 2D materials [41], heterostructures, and related quantum materials [42]. Although IPSs are less prone to electron-phonon scattering, the high energy and spatial resolution of FER spectroscopy in an STM setup opens the possibility of probing electron-phonon coupling physics of borophene polymorphs at the nanoscale, which is critical to understanding the predicted phonon-mediated superconductivity in borophene [20].

**Acknowledgements**

X.L. and M.C.H. acknowledge support from the Office of Naval Research (ONR N00014-17-1-2993) and the National Science Foundation Materials Research Science and Engineering Center (NSF DMR-1720139). L.W. and B.I.Y. acknowledge support from the Electronics Division of the U.S. Army Research Office (W911NF-16-1-0255) and the Robert Welch Foundation (C-1590).**References**

[1] A. J. Mannix, B. Kiraly, M. C. Hersam, and N. P. Guisinger, *Nat. Rev. Chem*. **1**, 0014, (2017).
[2] X. Liu and M. C. Hersam, *Adv. Mater*. **30**, 1801586 (2018).
[3] G. Binnig, K. H. Frank, H. Fuchs, N. Garcia, B. Reihl, H. Rohrer, F. Salvan, and A. R. Williams, *Phys. Rev. Lett.* **55**, 991 (1985).
[4] M. Winter, E. V. Chulkov, and U. Höfer, *Phys. Rev. Lett.* **107**, 236801 (2011).
[5] Z. Zhang, Y. Yang, G. Gao, and B. I. Yakobson, *Angew. Chem. Int. Ed.* **54**, 13022 (2015).
[6] G. Giovannetti, P. A. Khomyakov, G. Brocks, V. M. Karpan, J. van den Brink, and P. J. Kelly, *Phys. Rev. Lett.* **101**, 026803 (2008).
[7] P. A. Khomyakov, G. Giovannetti, P. C. Rusu, G. Brocks, J. van den Brink, and P. J. Kelly, *Phys. Rev. B* **79**, 195425 (2009).
[8] D. Jose and A. Datta, *J. Phys. Chem. C* **116**, 24639 (2012).
[9] M. E. Dávila and G. Le Lay, Sci. Rep. **6**, 20714 (2016).
[10] A. J. Mannix, X.-F. Zhou, B. Kiraly, J. D. Wood, D. Alducin, B. D. Myers, X. Liu, B. L. Fisher, U. Santiago, J. R. Guest, M. J. Yacaman, A. Ponce, A. R. Oganov, M. C. Hersam, and N. P. Guisinger, *Science* **350**, 1513 (2015).
[11] L. Zhu, B. Zhao, T. Zhang, G. Chen, and S. A. Yang, *J. Phys. Chem. C* **123**, 14858 (2019).
[12] J. Deng, B. Xia, X. Ma, H. Chen, H. Shan, X. Zhai, B. Li, A. Zhao, Y. Xu, W. Duan, S.-C. Zhang, B. Wang, and J. G. Hou, *Nat. Mater.* **17**, 1081 (2018).
[13] F. Schulz, R. Drost, S. K. Hämäläinen, T. Demonchaux, A. P. Seitsonen, and P. Liljeroth, *Phys. Rev. B* **89**, 235429 (2014).
[14] F. Craes, S. Runte, J. Klinkhammer, M. Kralj, T. Michely, and C. Busse, *Phys. Rev. Lett.* **111**, 056804 (2013).
[15] B. Feng, J. Zhang, Q. Zhong, W. Li, S. Li, H. Li, P. Cheng, S. Meng, L. Chen, and K. Wu, *Nat. Chem.* **8**, 563 (2016).
[16] X. Liu, Z. Zhang, L. Wang, B. I. Yakobson, and M. C. Hersam, *Nat. Mater.* **17**, 783 (2018).
[17] X. Liu and M. C. Hersam, *Sci. Adv.* **5**, eaax6444 (2019).
[18] X. Liu, L. Wang, S. Li, M. S. Rahn, B. I. Yakobson, and M. C. Hersam, *Nat. Commun.* **10**, 1642 (2019).
[19] B. Feng, O. Sugino, R.-Y. Liu, J. Zhang, R. Yukawa, M. Kawamura, T. Iimori, H. Kim, Y. Hasegawa, H. Li, L. Chen, K. Wu, H. Kumigashira, F. Komori, T.-C. Chiang, S. Meng, and I. Matsuda, *Phys. Rev. Lett.* **118**, 096401 (2017).
[20] M. Gao, Q.-Z. Li, X.-W. Yan, and J. Wang, *Phys. Rev. B* **95**, 024505 (2017).
[21] N. Gonzalez Szwacki, A. Sadrzadeh, and B. I. Yakobson, *Phys. Rev. Lett*. **98**, 166804 (2007).
[22] H. Tang and S. Ismail-Beigi, *Phys. Rev. B* **80**, 134113 (2009).
[23] H. Tang and S. Ismail-Beigi, *Phys. Rev. Lett.* **99**, 115501 (2007).
[24] E. S. Penev, S. Bhowmick, A. Sadrzadeh, and B. I. Yakobson, *Nano Lett.* **12**, 2441 (2012).
[25] S.-Y. Xie, X.-B. Li, W. Q. Tian, N.-K. Chen, X.-L. Zhang, Y. Wang, S. Zhang, and H.-B. Sun, *Phys. Rev. B* **90**, 035447 (2014).
8

**Figures**

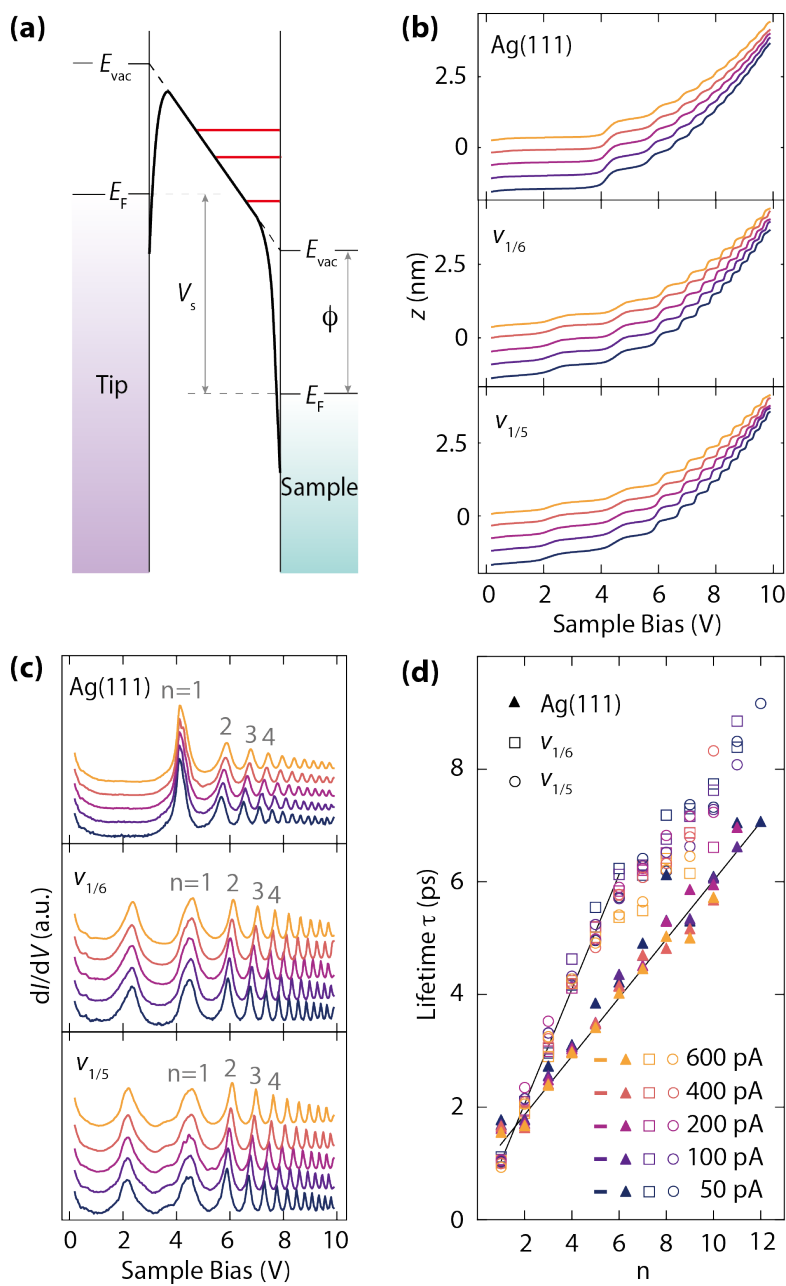

**Figure 1**. Image potential states of borophene polymorphs and Ag(111). (a) Schematic of the band alignment in the Fowler–Nordheim tunneling regime with Stark-shifted IPSs. $E_{vac}$: vacuum level, $E_F$: Fermi level, $V_s$: sample bias. (b) $z$-$V$ spectroscopy on Ag(111), $v_{1/6}$ borophene, and $v_{1/5}$ borophene at different current setpoints specified in the legend of (d). (c) Simultaneously acquired $dI/dV$-$V$ spectra showing IPSs peaks at current setpoints specified in the legend of (d). (d) Measured lifetimes of the image potential states shown in (c). The black lines are a guide to the eye.

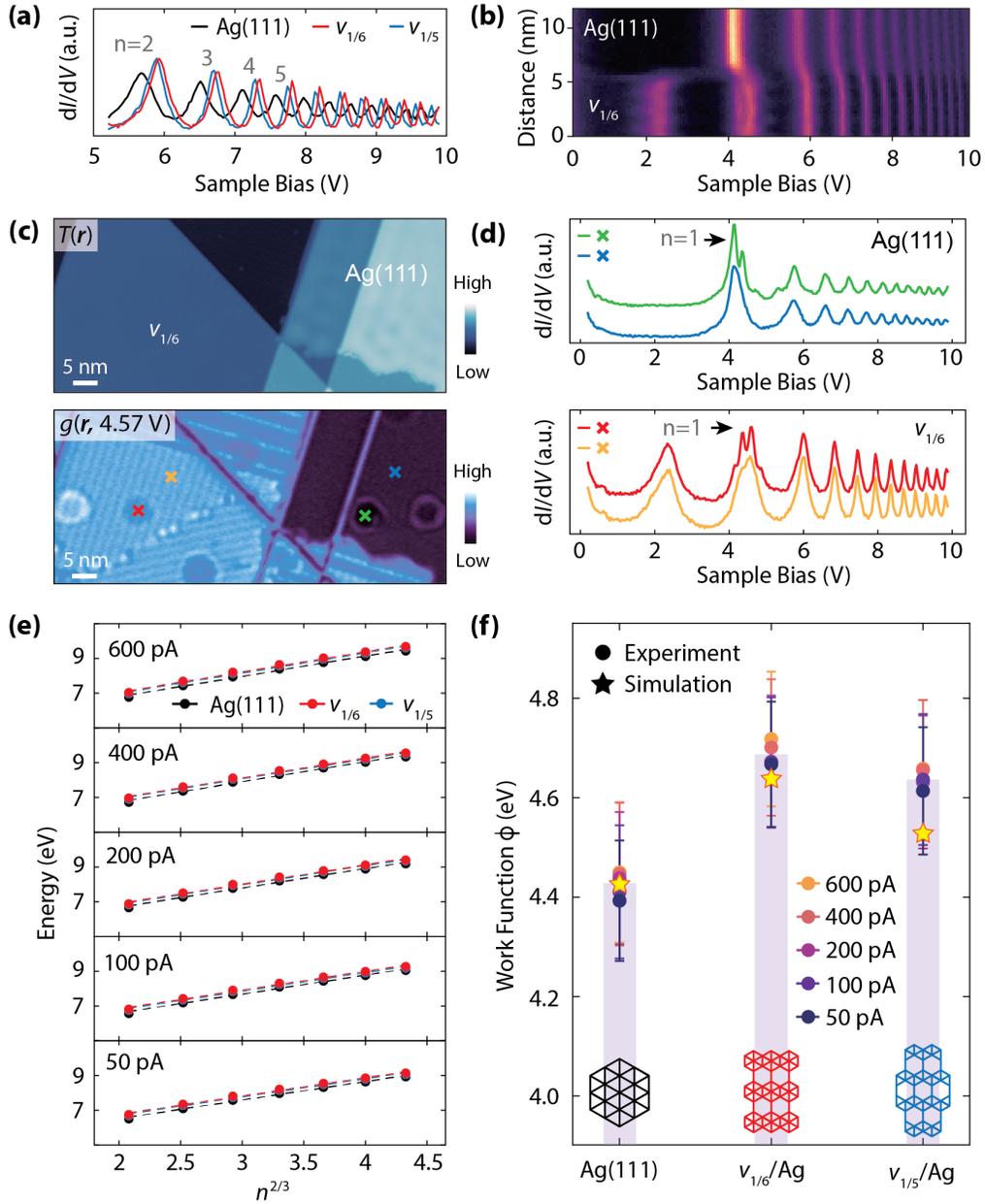

**Figure 2**. Local work function characterization. (a) Comparison of the IPSs peaks on Ag(111), $v_{1/6}$ borophene, and $v_{1/5}$ borophene. (b) A series of d$I$/d$V$-$V$ spectra taken across the interface between Ag(111) and $v_{1/6}$ borophene. (c) Topography ($T(\boldsymbol{r})$, top) and differential conductance ($g(\boldsymbol{r})$, bottom) map at the energy of $n = 1$ IPS of the same area. $V_s$ = 10 mV, $I_t$ = 300 pA for $T(\boldsymbol{r})$. (d) FER spectra taken at the accordingly colored crosses in (c). (e) Extracted and fitted IPS peak positions. (f) Comparison of experimentally obtained and DFT calculated LWFs of Ag(111), $v_{1/5}$ borophene, and $v_{1/6}$ borophene.

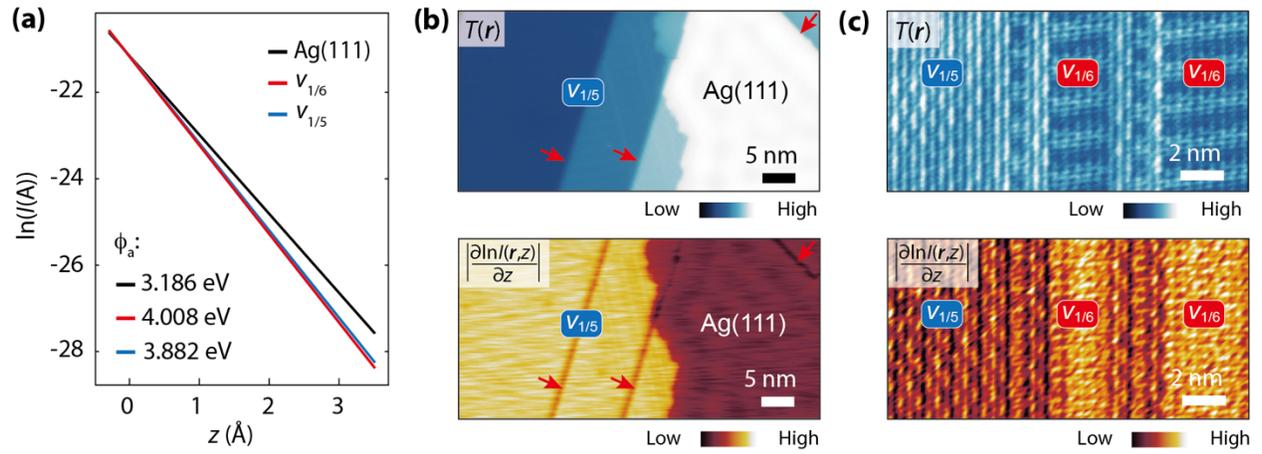

**Figure 3**. Apparent barrier height measurements. (a) ABHs measured on Ag(111), $v_{1/6}$ borophene, and $v_{1/5}$ borophene, revealing larger ABHs for borophene polymorphs compared to Ag(111). (b,c) Topography ($T(\mathbf{r})$) and simultaneous $|\frac{\partial \ln I(\mathbf{r},z)}{\partial z}|$ images on (b) the boundary between $v_{1/5}$ borophene and Ag(111) ($V_s = -20$ mV, $I_t = 500$ pA) and (c) the boundary between $v_{1/5}$ borophene and $v_{1/6}$ borophene ($V_s = -72$ mV, $I_t = 530$ pA).

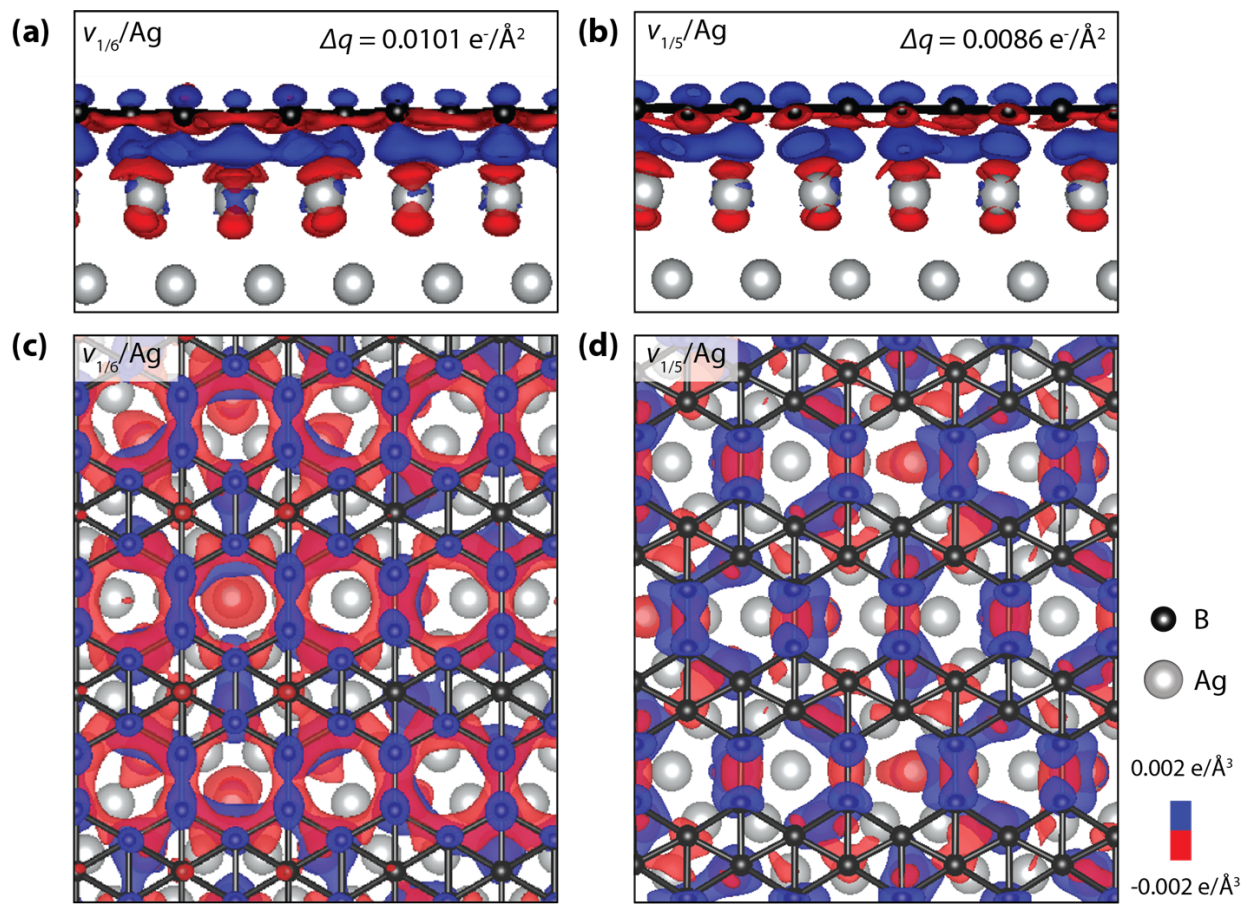

**Figure 4**. Charge transfer between borophene polymorphs and the Ag(111) substrate as determined from density functional theory calculations. (a,b) Cross-sectional view of the charge redistribution in (a) $v_{1/6}$ borophene and (b) $v_{1/5}$ borophene on Ag(111). (c,d) In-plane view of the charge redistribution in (c) $v_{1/6}$ borophene and (d) $v_{1/5}$ borophene on Ag(111).

13